\def\nall{1303}        
\def\ncomb{498}        
\def\nstarfields{55}    
\def\ngalfields{122}    

\documentclass[printer]{aa}
\usepackage{graphics}
\begin{document}
\title{Cosmic Shear from STIS Pure Parallels}
\authorrunning{Pirzkal et al.}
\subtitle{I Data}

\author{N.~Pirzkal\inst{1}, L.~Collodel\inst{1}, T.~Erben\inst{3,}\inst{4,}\inst{7}, R.~A.~E.~Fosbury\inst{1}, W.~Freudling\inst{1}, H.~H\"ammerle\inst{2,}\inst{3}, B.~Jain\inst{5}, A.~Micol\inst{1}, J.-M.~Miralles\inst{1,}\inst{2}, P.~Schneider\inst{2,}\inst{3}, S.~Seitz\inst{6}, S.~ D.~ M.~ White\inst{3}}

\offprints{N.~Pirzkal \email{npirzkal@eso.org} }

\date{February 19, 2001; accepted for publication in A\&A}
\institute{ST-ECF, Karl-Schwarzschild Str. 2, Garching bei M\"unchen D-85748, Germany
\and
Institut f\"ur Astrophysik und Extraterrestrische Forschung der
Universit\"at Bonn, Auf dem H\"ugel 71, D-53121 Bonn, Germany
\and
Max-Planck-Institut f\"ur Astrophysik, Karl-Schwarzschild Str. 1, D-85741 Garching, Germany
\and
Institut d'Astrophysique de Paris, 98bis Boulevard Arago, F-75014 Paris, France
\and
Department of Physics and Astronomy, University of Pennsylvania
        209 S. 33rd Street, Philadelphia, PA 19104 USA
\and
Universit\"ats-Sternwarte M\"unchen, Scheiner Str. 1, D-81679 M\"unchen, Germany
\and
Observatoire de Paris, DEMIRM 61, Avenue de l'Observatoire, F-75014 Paris
}

\abstract{
Following the second HST servicing mission in 1997 when the STIS
instrument was installed and the capability for parallel observations
was enhanced, a substantial archive of non-proprietary parallel data
has been accumulating. In this paper, we discuss the use of unfiltered
STIS imaging data for a project that requires deep observations along as
many independent lines-of-sight as possible. We have developed a
technique to determine which datasets in the archive can safely be
co-added together and have developed an iterative co-addition technique
which enabled us to produce \ncomb\ high-quality, deep images. The
principal motivation for this work is to measure the Cosmic Shear on
small angular scales and a value derived from these data will be
presented in a subsequent paper. A valuable by-product of this work is
a set of high quality combined fields which can be used for other
projects. The data are publicly available at
http://www.stecf.org/projects/shear/  
}
\maketitle

\section{Introduction}

In June 1997, parallel observations using the Space Telescope Imaging
Spectrograph (STIS) on the Hubble Space Telescope started to be taken in
substantial numbers as the STIS Parallel Survey (SPS) was carried out.
The result of the SPS was to generate thousands of relatively short
exposure images and slitless spectra taken along many different
lines-of-sight. The data were non-proprietary and were made available
almost immediately.

We are using the imaging data to investigate the distortion of
background galaxies by the gravitational field of the large scale
matter distribution in the Universe, also known as Cosmic Shear
(Mellier \cite{mellier1999}; Bartelmann \& Schneider
\cite{bartelmann2001}). This effect was recently detected from the
ground (Van Waerbeke et al. \cite{waerbek}; Bacon et
al. \cite{bacon2000}; Kaiser et al. \cite{kaiser2000}; Maoli et
al. \cite{maoli2000}; Wittman et al. \cite{wittman2000}; Van Waerbeke
et al. \cite{waerbek2001}) and from space (Rhodes et
al. \cite{rhodes2000}). The typical object sizes that have to be
measured are on the order of the seeing size of typical ground-based
observations. Such observations are therefore better carried out using
the stable, high spatial resolution that is provided by space-based
cameras such as those on HST. Due to intrinsic cosmic variance, this
project requires many observations of separate, independent fields
along different lines-of-sight, each containing several tens of faint,
small background galaxies from which high precision measurements of
shapes can be made.

This paper, the first in a series, describes the nature of the
available STIS data (Sects. \ref{sec:STIS} and \ref{sec:SPS}), the
selection of the input images from the pool of existing archived
images (Sect. \ref{sec:selection}), and the data reduction that has
been performed (Sect. \ref{sec:dataredux}) to produce \ncomb\ deep,
co-added fields. We conclude with a description of the extended object
content of the reduced data (Sect. \ref{sec:galdescription}).  A
subsequent paper will demonstrate that the properties of the STIS
imaging data, the SPS data in particular, make them very appropriate
for an investigation of the Cosmic Shear on small scales, and will
present Cosmic Shear measurements derived from these data. All of the
co-added images discussed in this paper are publicly available at
http://www.stecf.org/projects/shear/.

\section{Data}
\label{sec:data}
\subsection{STIS}
\label{sec:STIS}
STIS CCD images provide good depth, excellent resolution and adequate
sampling of the telescope point spread function (PSF), making them a
prime choice for our Cosmic Shear study.  The STIS CCD is
sensitive to wavelengths ranging from 2500 to 11000 $\AA$ and its
field of view is $51\arcsec \times 51\arcsec$.  The STIS CCD pixel
size is $0.05\arcsec$ and is the result of a compromise between
properly sampling the STIS CCD PSF (FWHM of $0.05\arcsec$ at 500nm)
and having a field of view which is as large as possible. The full
imaging field of the STIS CCD can only be used in an unfiltered mode
but this results in a high sensitivity. In this mode, called the CLEAR
filter mode, the attained throughput is significantly higher, and the
bandpass broader, than with other HST instruments such as WFPC2. The
limiting magnitude in a 1 hour exposure, for a A0V star, at a
signal-to-noise ratio of 5, reached in the STIS CLEAR filter mode, is
$V_{AB}=28$ while WFPC2 combined with the broad F606W filter only
reaches a limiting magnitude of $V_{AB}=27.2$ (Gregg \& Minniti
\cite{stisccd}). These numbers show how much more efficient the STIS
CLEAR filter mode is at capturing photons. The very broad bandpass
does, however, bring some disadvantages, the principal ones being the
non-standard photometric band and the substantial variation of the
diffraction pattern across the band which results in an effective PSF
which depends on object colour. As we will see in Sect.
\ref{sec:galdescription}, in 1 orbit with the CLEAR filter, STIS can
image a  sufficient number (20--30)  of field  galaxies for  a  Cosmic
Shear measurement to   be attempted. More  detailed  information about
STIS and its performance can be found  in the STIS Instrument Handbook
(Leitherer et  al.   \cite{stishandbook}), and in Gregg \& Minniti
(\cite{stisccd}),  Kimble  et  al.  (\cite{kimble}),  and Gilliland et
al. (\cite{Gilliland}).

\subsection{STIS Parallel Survey (SPS)}
\label{sec:SPS}
The HST observing mode called the Parallel Mode became widely available
in June 1997 following the installation of solid state recorders on-board
HST during the second servicing mission. In Parallel Mode, more than one
instrument can be used at a time: the primary instrument carries out its
scheduled science observations while other instruments, the parallel
instruments, are also allowed to collect data. The pointings of the
parallel instruments are dictated by the pointing of the primary
instrument and by the location of each instrument in the telescope focal plane.
Parallel observations are implemented on a strict basis of
non-interference with the primary instrument, imposing strong
restrictions on the exposure times and dithering patterns of parallel
observations. When parallel observations are not specifically requested
by the user of the primary instrument, `Pure Parallel' observations are
carried out which do not have a proprietary period and are made public
immediately after they have been taken.

The SPS observing program (e.g., Gardner et al. \cite{gardner98}) was
created in 1997. Its goal was to make use of the Pure Parallel Mode
opportunities of STIS by taking pure parallel direct and grism
observations in an automated, scripted manner any time that another
instrument was used and STIS remained available. The actual
implementation of the SPS has been under the supervision of the
Parallel Survey Working Group at STScI.

Some of the SPS data were taken in the CLEAR filter mode, had a
relatively short integration time of a few hundred seconds to minimize
the impact of cosmic rays (Smette \& Hill \cite{smette}), and were
taken in what is called CR-SPLIT mode. This mode consists of taking at
least two successive images in order to later facilitate the removal
of cosmic ray impacts from each image. It also ensures that telescope
tracking errors (Sect. \ref{sec:tracking} and Appendix) do not
accumulate over long periods of time. During the period from June 1997
to October 1998 (date on which images stopped being taken in CR-SPLIT
mode), we found the SPS proposals number 7781, 7783, 7908, 7910, 7911,
8062, 8064, and 8084 to contain datasets that met our
requirements. The number of datasets available is summarized in Table
\ref{tab:SPS}. A subsequent Cycle 9 GO program (8562; PI Schneider) is
continuing the STIS parallel imaging observations in a non-proprietary
mode: the analysis of these data is not included here but will be
discussed in another paper.

\begin{table}
     \caption[]{Content of the SPS proposals listed in Sect.
     \ref{sec:SPS}. Starting with a total of 2149 CR-SPLIT=2 images
     (not counting grism SPS exposures), 1312 images were identified
     to also be unbinned, and taken in the CLEAR filter. 1303 of these
     (9 images were rejected because of cosmetic defects) were used to
     produce \ncomb\ deep, long-exposure time, co-added images.}
     \label{tab:SPS}

      \[
           \begin{array}{p{.75\linewidth}ll}
            \hline
            \noalign{\smallskip}
            Type    & Number\\
            \noalign{\smallskip}
            \hline
		CR\_SPLIT datasets & 2149 \\
		CR\_SPLIT+Unbinned datasets& 1512 \\
		CR\_SPLIT+Unbinned+CLEAR datasets& 1312 \\
		Selected datasets & 1303 \\
		Co-added images & \ncomb\ \\
            \noalign{\smallskip}
            \noalign{\smallskip}
            \hline
         \end{array}
      \]

\end{table}

\subsection{Pointing and Tracking Precision of SPS}
\label{sec:tracking}

One should distinguish between two separate telescope pointing errors:
the first relates to where the telescope is actually pointing at the
time when an image was taken. The second is the ability of the
telescope system to keep the telescope pointing exactly at the desired
position during the course of an image acquisition. 

Having identified an appropriate set of images from the SPS, we
remained particularly concerned about the amount of image distortion
that could have been introduced by the relative instability of the
telescope tracking. There are in fact two distinct effects that can
result in less than perfect tracking conditions during an SPS
exposure. One is the inherent spacecraft jitter while the other one is
the problem of differential velocity aberration.  The effect of the
former is seemingly random and usually leads to a symmetrical widening
of the effective instrumental PSF. Differential velocity aberration,
on the other hand, introduces a systematic drift of the pointing of
parallel instruments (See Appendix A).

Random pointing errors are caused by a whole range of processes within
the spacecraft, ranging from thermally-driven jumps to minor cosmic
dust impacts. While the absolute accuracy of HST's pointing varies
from $0.5\arcsec$ to $2\arcsec$, relative astrometry between
successive images, taken during a single telescope visit and using the
same guide stars, is typically good down to between $0.001\arcsec$ and
$0.050\arcsec$.

The tracking errors caused by the differential velocity aberration
depend on the direction of the observed target relative to the
direction of the spacecraft velocity vector. While the pointing of the
primary instrument takes this effect into account, parallel
instruments are affected by a different amount of velocity aberration
since they are at a different position in the telescope focal
plane. The resulting un-corrected drift in a 400 second STIS exposure
can be as large as $0.005\arcsec$ (0.1 STIS CCD pixel). A more in-depth
discussion of the effect of differential velocity aberration is
presented in Appendix A of this paper.

While pointing and tracking errors can be expected during the course
of an exposure, one can easily check the quality of an HST observation
by using information obtained from the on-board Fine Guide Sensors
(FGS) system. Under good conditions, HST observations are obtained
while the positions of two guide stars are being monitored by the
FGS. This information feeds the closed-loop tracking system of the
telescope, but it is also recorded a few times a second and
incorporated into the Observation Log files which are made available
to users.  The position of the guide stars as a function of time can
be extracted from the Observation Log, allowing one to check how well
the telescope tracked during the course of an exposure. Given those
FGS guide star measurements, and given knowledge of focal plane
geometry and distortions, one can map any point in the telescope's
focal plane, such as an aperture, onto RA and Dec as a function of
time, during the course of an observation (``jitter ball'').  This was
done, for each exposure, and how well the telescope was tracking was
checked by determining the magnitude and asymmetry of the ``jitter
ball''.

\begin{figure} 
\resizebox{\hsize}{!}{\includegraphics{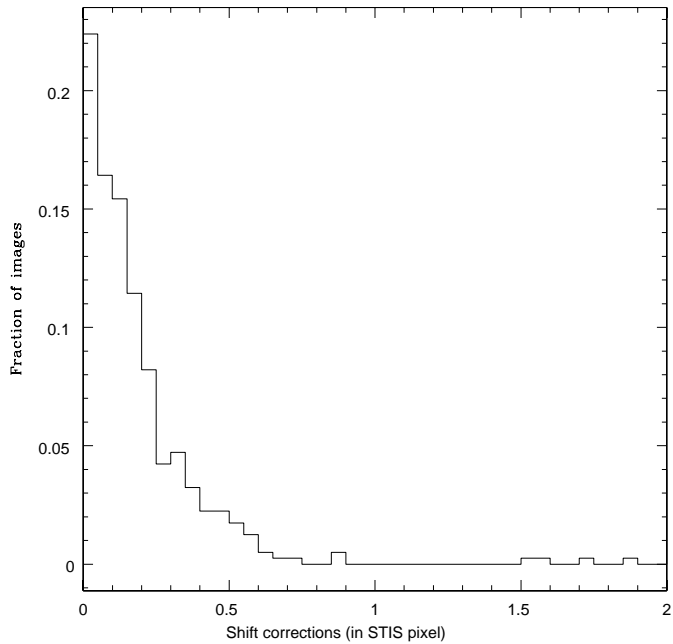}}
\caption{Histogram of the differences between the values of shifts computed 
using the available WCS information and the shifts computed using
jitter data. This histogram does not include 4$\%$
of the images for which the WCS-derived shifts were off by more than 2
pixels.
\label{np3}} 
\end{figure}

The jitter data can also be used to determine the accurate relative
pointing of a set of images, if those were taken in a single telescope
visit and using the same guide stars. This provides a good alternative
to simply using the available World Coordinate System (WCS)
information stored in the image headers since the latter is only a
record of where the telescope should have been pointing and does not
indicate where the telescope was actually pointing during the
observation (Micol et al. \cite{amicol1}). In Fig. \ref{np3}, we
present the result of using the jitter data by showing a histogram of
the differences between the WCS coordinates and the coordinates
derived using jitter data. We found that 50\% of the WCS-derived
offsets differed from the jitter-derived ones by more than 0.3 pixel,
while 4\% of them differed by more than 2 pixels.

\subsection{Selecting the SPS datasets} 
\label{sec:selection}

We restricted our analysis to a sub-set of the available SPS data
which satisfies the following conditions: An image had to be taken in
the CLEAR filter mode, in a CR-SPLIT mode, be unbinned (with a pixel
size of $0.050\arcsec$), and had to have an associated ``jitter ball'' rms
value smaller than $0.005\arcsec$ (0.1 STIS CCD pixel) (Sect.
\ref{sec:tracking}).  We then determined which images could be
combined together to produce deeper images. Starting with all of the
jitter data for the time period covered by the proposals we were
interested in (Sect. \ref{sec:SPS}), we selected images which were
taken consecutively during a single telescope visit using the same
telescope roll angle, and which were offset by no more than one
quarter of the field of view. Such groups of images were logically
associated into what we refer to as a STIS Association (Micol et al. \cite{amicol1})
and the relative offsets between the members of each STIS Association
were computed using the available jitter data. A histogram of the
exposure times of the
\nall\ individual CR-SPLIT=2 images that were selected is plotted in
Fig. \ref{fig:norsept99raw}.

\section{Reduction procedure}
\label{sec:dataredux}
Since a project such as ours can gain a lot by using deeper images in
which more field galaxies can be detected, our data reduction and
co-addition procedures were developed to co-add images together
without introducing any artificial distortion in the final co-added images.

\subsection{Basic Data reduction of individual SPS datasets}

All raw datasets were reduced using the latest version of the standard
STIS data reduction pipeline (CALSTIS) and using the best available
calibration files. As mentioned in Sect. \ref{sec:SPS}, only SPS
images taken using a CR-SPLIT=2 acquisition mode were used for this
project, which greatly helped the removal of cosmic ray impacts using
just CALSTIS.  CALSTIS was also used to bias subtract and flat-field
the raw images.  The use of the best available calibration files, such
as the STIS weekly darks, proved to be a good way to reduce the number
of un-corrected STIS hot pixels (pixels with larger than normal dark
currents and which affect about 2-3$\%$ of the pixels) by a factor of
about two, down to a number of about 1000.

\subsection{Additional Data Reduction}
\label{sec:additdatared}

Bad pixel maps, containing the location of un-corrected hot pixels,
cosmic rays, as well as the vertical streaks that are associated with
those were then created for each of the images processed by CALSTIS.

Hot pixels and residual cosmic ray impacts were located by flagging
any pixel which deviated by more than 8$\sigma$ from the surrounding
pixels. The group of surrounding pixels included all of the adjacent
pixels but excluded the pixels in the same column as the pixel being
examined since those were often affected by the cosmic ray impact.

Noisy, one-pixel-wide columns (2 to 10 per image) were detected by
examining groups of 50 vertically adjacent pixels at a time. When the
median of such a column segment was found to be 1$\sigma$ higher
than the corresponding column segments in both the preceding and
following columns, that particular section, as well as all the pixels
below, were flagged as bad. This was shown to accurately detect $95\%$
of these noisy columns.

No further cleaning or flagging of cosmetic defects in the images,
such as spikes and ghosts caused by bright stars (Leitherer et
al. \cite{stishandbook}) was performed at this stage.
 
\subsection{Image co-addition} 
\label{sec:coadding}
Once individual images were reduced and their bad pixel maps were
generated, the actual process of co-adding the members of a STIS
Association was performed using an iterative cross-correlation and
Drizzle (Fruchter \& Hook \cite{fruchter}) technique which was
specifically developed for this purpose.

The use of Drizzle as the core of this iterative process allowed us to
properly exclude bad pixels using the image masks generated earlier
(Sect. \ref{sec:additdatared}), to apply sub-pixel shifts, to
incorporate the known STIS distortion map (Goudfrooij \cite{ISR97}) by
un-distorting individual images first, and to project the combined
drizzled image onto a finer, sub-sampled grid. A sub-sampling of a
factor of two was used in order to reduce the spatial smearing
produced when small under-sampled objects are shifted by a fraction of
a pixel.

While the offsets derived using the jitter data were more accurate
than WCS-derived ones, the use of cross-correlation allowed us to
further refine our estimates of the relative offsets between STIS
Association members.  The method, which is iterative, proceeds as
follows: Using the initial offsets determined from the available
jitter data, a first estimate of the combined image was generated
using the Drizzle IRAF routine. In addition, similarly shifted and
drizzled versions of each individual STIS Association member were also
produced. SExtractor (Bertin \& Arnouts \cite{arnouts}) was then used
to identify objects in both the combined image and the individually
drizzled STIS Association members and to produce background-free
versions of these images, where background regions were set to a value
of zero. The individually drizzled, background-free STIS Association
members were then separately cross-correlated with the background-free
combined image to determine a set of shift corrections. These were
then added to the STIS Association member offsets, and the entire
process was repeated using the new offsets estimates until the method
converged and the computed shift corrections became smaller than 0.05
pixels. This method usually converged in 4 to 10 iterations.

\subsection{Resulting output}

Starting with the original pool of datasets from the proposals listed in
Sect. \ref{sec:SPS}, \nall\ individual CR-SPLIT=2, unbinned, CLEAR
filter STIS images were combined together to produce \ncomb\ co-added STIS
pure parallel Associations.

\begin{figure}
\resizebox{\hsize}{!}{\includegraphics{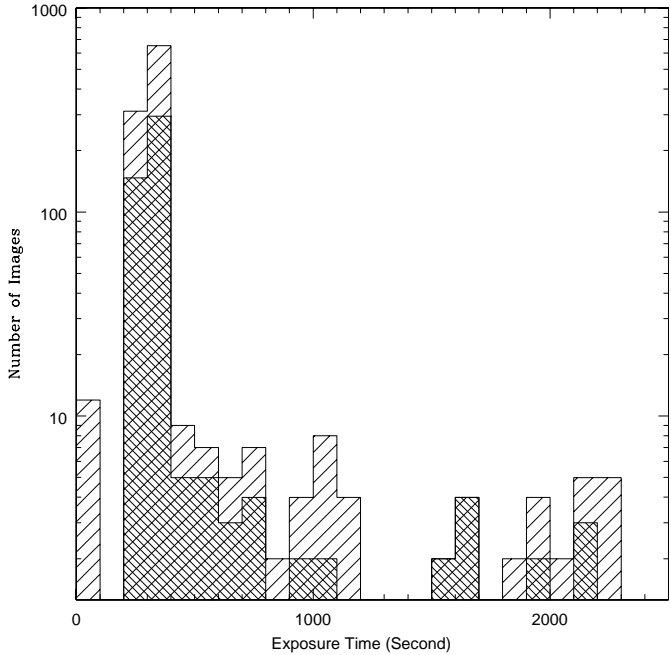}}
\caption{Histogram of the exposure times of the individual STIS Pure
Parallel images (coarse hashes).  Each image is made up of two
separate exposures, each with half of the exposure time (CD-SPLIT=2
mode). The subset of images with galactic latitude greater than 30
degrees is also plotted (tight hashes).
\label{fig:norsept99raw} } 
\end{figure}

\begin{figure}
\resizebox{\hsize}{!}{\includegraphics{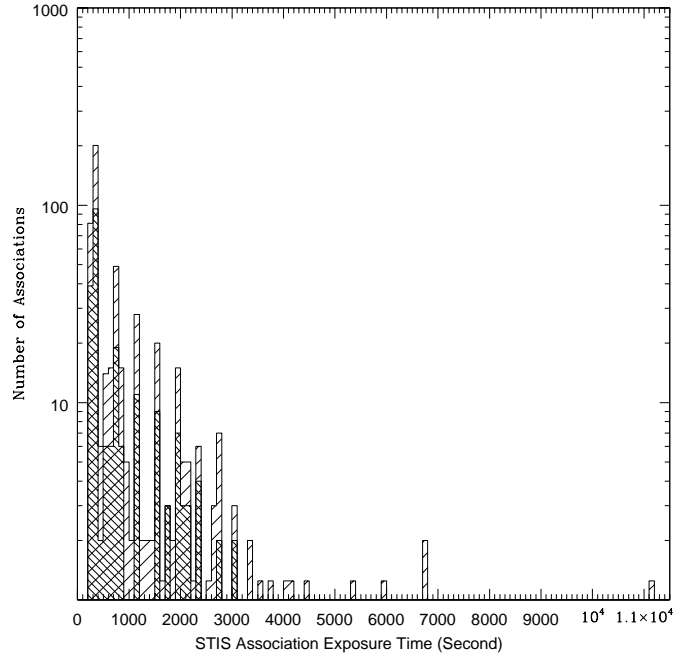}}
\caption{Our STIS Associations effective exposure times (coarse
hashes).  Associations with galactic latitude larger than 30 degrees
are also plotted (tight hashes).
\label{fig:norsept99red}} 
\end{figure}

The effective integration time of the co-added images is a function of
the number of members in each STIS Association and of the exposure
time of the individual Association members. In
Fig. \ref{fig:norsept99raw}, we plot the distribution of exposure
times of the individual STIS Association members. The typical
offsets between individual members of a STIS Association was 12 STIS
CCD pixels. A histogram of the effective exposure
times of the combined STIS Associations images is presented in
Fig. \ref{fig:norsept99red}.

\subsection{The applied offsets} 

We found that about 4$\%$ of our images had recorded WCS values that
were off from the jitter-derived coordinates by more than $0.1\arcsec$ (2
STIS CCD pixels). Such large offset errors could not have been recovered
using our iterative cross-correlation technique which required a
reasonably good first estimate of the offsets to start with. Extensive
checking of the accuracy of these final images offsets was done and is
summarized in the next Sect. \ref{sec:imtechn}.

\section{Testing of the co-addition procedure}
\label{sec:imtechn}

The reduction and image co-addition procedure outlined in Sect.
\ref{sec:coadding} was thoroughly tested by applying the technique
to a set of simulated STIS Associations.

We created simulated STIS Associations containing objects with
positions, magnitudes, shapes, and spatial distributions realistically
modeled after the content of one of our high galactic latitude SPS
field containing 75 objects. We used Tiny
Tim\footnote{http://www.stsci.edu/software/tinytim/tinytim.html} to
generate the STIS PSF which we used to produce our simulated images.

As a test, 20 simulated STIS Associations, each containing 6 dithered
member images, were constructed. In each of these simulated
Associations, we varied the actual offsets between each Association
member, assigning them realistic values. Poisson noise, background
sky, and a realistic number of cosmic ray impacts were added using
the IRAF tasks MKOBJECTS and MKNOISE. These simulations were processed
using the procedure outlined in Sects. \ref{sec:additdatared} though
\ref{sec:coadding}. In every single one of our 20 simulations we were
able to re-align the individual STIS Association members to within
$0.001\arcsec$ (0.02 STIS CCD pixel).

This test was then repeated a second time using a new set of 20
simulated STIS Associations. This time however the 14 brightest
objects amongst the 75 objects used initially were removed. We also
generated Associations with different S/N by arbitrarily increasing
the magnitudes of the 61 objects in our simulated fields, up to 0.7
magnitude per object.  Some Associations were also generated with and
some without cosmic rays or hot pixels to verify that our cosmic rays
and hot pixel removal method (see Sect. \ref{sec:additdatared}) did
not influence our offset determinations. Our co-addition method proved
very efficient, and the individual STIS Association member offsets
were re-constructed. Even in the case of simulations with objects
having a signal to noise ratio which was half of the value measured on
the observed image, we were able to recover the proper offsets to
within $0.004\arcsec$ (0.08 STIS CCD pixel).

We additionally tested the sensitivity of our method to the
object-blurring effect that the STIS Charge Transfer Inefficiency
(CTI) might cause as the centroid of objects located at low Y
coordinates is progressively shifted to lower Y coordinates relative
to object higher up in the CCD.  We tested this by selecting one of
our STIS associations containing a large number of members (14 members
of 300s each, with relative offsets of as much as 40 pixels), and
determined the shifts separately for the top and bottom halves of the
images. This test was made once using only the brighter (10 per half
field), and once more using only the fainter objects (5 connected STIS
CCD pixels 3$\sigma$ above the background). In all cases, the derived
offsets were consistent (to within $1/20$ of a STIS CCD pixel) with
the offsets derived using all the objects in the entire image. This
shows that the STIS CTI does not affect our ability to properly
determine the offsets between STIS association members.  The test
using only bright objects additionally shows that our image cross
correlation is not significantly affected by the presence of STIS PSF
ghosts, which previously described tests using Tiny Tim could not
include.  

While flux conservation was also tested, we do not present the test
results in this paper because they only confirm the known ability of
the Drizzle algorithm to conserve flux (Fruchter \& Hook
\cite{fruchter}). We confirmed through these tests that our method
preserves the shape of individual objects. An extensive description of
these results will be presented in the second paper of this series.

\section{Sky coverage, number counts, and galaxy sizes}
\label{sec:galdescription}

\begin{figure}  
\resizebox{\hsize}{!}{\includegraphics{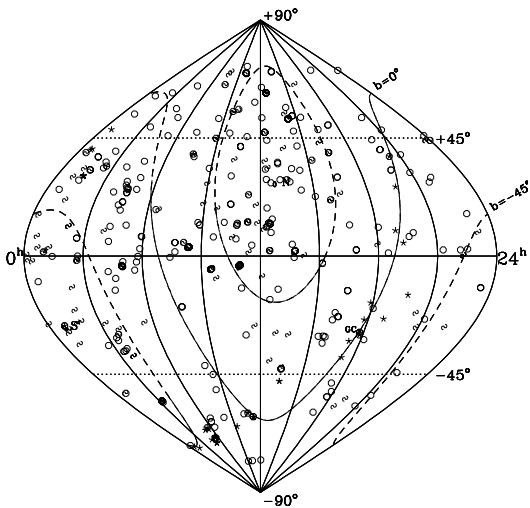}}
\caption{Galactic coordinates of the \ncomb\ co-added fields. The stars, swirls, and empty circles represent stellar fields, galaxy fields, and non-classified fields respectively (See Sect. \ref{sec:numcount}) \label{fig:lball}}
\end{figure}

\subsection{Sky distribution, exposure times and magnitude limits }

The \ncomb\ fields that we produced from the STIS SPS data are spread
around the sky. Fig. \ref{fig:lball} shows the spatial distribution
of our datasets.

Our \ncomb\ co-added images have a very heterogeneous distribution of
exposure times, and the limiting magnitude reached on a field depends
on the number of images present in the STIS Association and on the
integration time of the individual images. The final distribution of
exposure times is plotted in Fig. \ref{fig:norsept99red}. Using
SExtractor, we estimate our limiting magnitude, for a 5 pixel
detection and for a signal to noise of 3.0, to be
$M_{AB}(CLEAR)=28.5$ in a 1 hour exposure.

\subsection{Galaxy number counts}
\label{sec:numcount}
Our final co-added images are distributed all over the sky
and are likely to be affected by various amounts of galactic
absorption. While this implies our sample of images is not
adequate for a detailed study of galaxy number counts, number counts
provide an interesting description of our data.

\begin{figure}
\resizebox{\hsize}{!}{\includegraphics{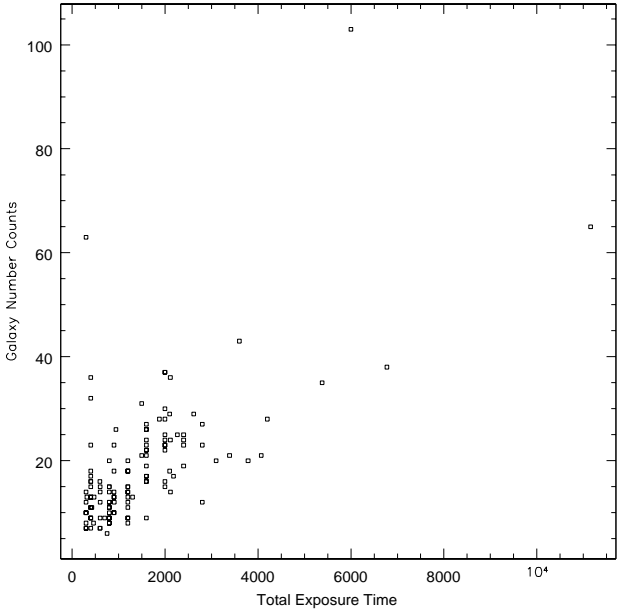}}
\caption{
Number of detected galaxies per STIS SPS Association plotted as a
function of exposure time, for the \ngalfields\ galaxy fields.
\label{fig:galnumcount}} 
\end{figure}

All of the co-added Associations were visually inspected and
classified as being possibly a member of two possible types of fields:
galaxy or stellar fields. Galaxy fields are fields where more than 10
extended objects were detected with the procedure described below while
stellar fields are fields where more than 100 point like objects, uniformly
distributed over the field, could be detected. Starting with the
\ncomb\ co-added STIS Associations, \nstarfields\ stellar fields and
\ngalfields\ galaxy fields were identified. SExtractor was used to
generate an object catalog of the galaxy fields. The main
detection criteria used during the extraction process were that each
object had to have a minimum of 5 pixels at least 3$\sigma$ brighter
than the local background. A Gauss\_4.0\_7x7 convolution kernel was
used. The complete SExtractor parameter file is being made available
at http://www.stecf.org/project/shear/. As seen in Fig.
\ref{fig:galnumcount}, the number of detected galaxies per co-added
image rises steadily from $12\pm 6$ to $29 \pm 12$ for total exposure
times ranging from 500 to 2500 seconds. The galaxy number counts then
rises more slowly as the exposure time increases past 2500
seconds. This flattening trend is caused by the intrinsic flattening
of the galaxy number counts at increased magnitude and because distant 
(and/or faint) objects
quickly become too small to be resolved by STIS (see Fig.
\ref{fig:galsize} and Fig. 1 in Gardner \& Satyapal \cite{gardner2000}) and to be
classified as galaxies. This leads to the conclusion that the optimal
integration time for a project using STIS which requires as many field
galaxies in as many independent lines of sight as possible is
somewhere between 2000 to 2500 seconds, which corresponds to
approximately 29 galaxies per field.

It should be noted that the redshift distribution of these objects,
which will be needed for the quantitative interpretation of our Cosmic
Shear measurement, can not be derived using solely the STIS SPS
data. This information will however be determined using photometric
redshifts obtained from lower resolution, deep, multicolor
VLT images.

\subsection{Galaxy sizes}
\begin{figure}
\resizebox{\hsize}{!}{\includegraphics{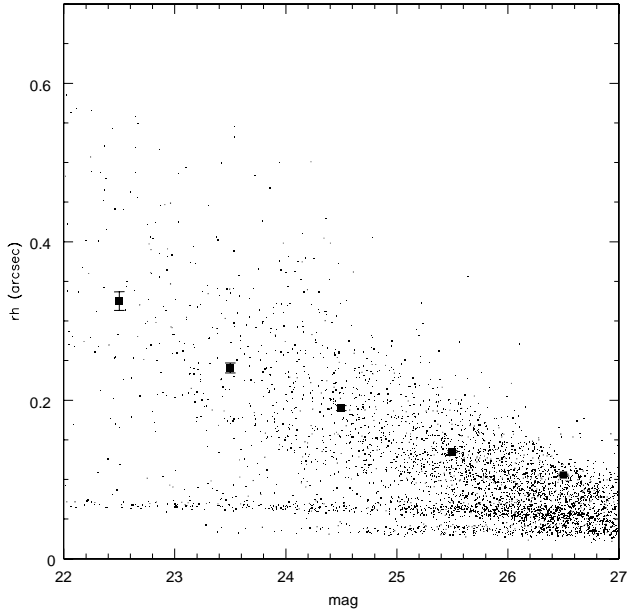}}
\caption{
Measured half-light radius of objects in our co-added STIS SPS
images. The horizontal distribution of points at a half-light radius
between $0.05\arcsec$ and $0.08\arcsec$ are caused by stars and other
un-resolved objects. The lowermost horizontal distribution of
points with a half-light radius ranging from $0.025\arcsec$ to
$0.030\arcsec$ is caused by spurious detections of single noisy pixels
by SExtractor. The average sizes of galaxies per magnitude bin are
also indicated. The filled squares mark the average sizes of objects
at each magnitude, counting only those with half-light radius $>
0.08\arcsec$. The error bars represent the $1\sigma$ level in the
error of the mean.  }
\label{fig:galsize} 

\end{figure}

While our co-added STIS SPS Associations provide deep, high resolution
images, it can be seen in Fig. \ref{fig:galsize} that the average
size of galaxies decreases as object CLEAR filter AB magnitudes
increase. We estimate that objects with magnitudes ranging from 25.5
to 26.5 have average sizes ranging from $0.14\arcsec$ to $0.11\arcsec$. This result
is consistent with the results published by Gardner \& Satyapal
(\cite{gardner2000}) which was based on 115 STIS observations of the
HDF-S, and with results from Odewahn et al. (\cite{odewahn1996}). From
our SPS data, one can expect most galaxies with magnitude ranging from
22 to 26 to have half-light radii ranging from $0.3\arcsec$ to $0.1\arcsec$ (6 to 2
STIS CCD pixels). The upper right portion of Fig. \ref{fig:galsize} is
however expected to suffer from incompleteness since it is likely to
be affected by our relative insensitivity to low surface brightness
objects.

\section{Conclusion}
We have begun an investigation of Cosmic Shear which requires many
images at different pointings containing a high number of faint
field galaxies. One source of such data has been provided by the SPS
program which has been producing a considerable amount of data since
June 1997. We have successfully accumulated the SPS data
taken from June 1997 to October 1998 and have used HST jitter data to
co-add as many of these images as possible. In this way, we have
produced a set of \ncomb\ co-added images which have individual
integration times ranging from 400 to 11000 seconds. The images were
co-added using an iterative cross correlation technique that we have
developed and which was shown using simulations to be able to
consistently match images to within 0.05 STIS CCD pixel.

We selected a set of \ngalfields\ images which contains a sufficient
number (10 to 30) of extended objects to make a meaningful Cosmic
Shear measurement.  The number counts of galaxies in these images are
consistent with previously published numbers (Gardner \& Satyapal
\cite{gardner2000}), and the optimal STIS integration time is between
2000 and 2500 seconds, leading to an average of 29 galaxies to be
detected at the $3\sigma$ level. The average sizes of the galaxies
with magnitude ranging from 22 to 26 vary from $0.3\arcsec$ to $0.1\arcsec$ and are
also consistent with previous observations (Gardner \& Satyapal
\cite{gardner2000};Odewahn et al. \cite{odewahn1996}).

The STIS galaxies have presumably a wide redshift distribution, which
is essentially unknown at this point. In order to transform a Cosmic
Shear measurement on the STIS images into a cosmological constraint,
we are currently carrying out lower resolution, deep VLT images to
determine the photometric redshift distribution of the STIS galaxies.

The analysis of these fields will be presented in Paper II. The
co-added data described in this paper is made available at
http://www.stecf.org/projects/shear/.

\begin{acknowledgements} We thank the staff of the STScI for their
hard work which, as a response to suggestions made by the Parallel
Working Group in 1997, enabled these observations to be carried
out. We also thank the two referees for their helpful comments. This
work has been partly supported at the ST-ECF by the ESO Director's
Discretionary Fund, the TMR Network ``Gravitational lensing: new
constraints on cosmology and the distribution of dark matter'' of the
European Community, and by a grant from the Verbundforschung
Astronomie/Astrophysik.  
\end{acknowledgements}

\appendix
\section{Differential velocity aberration}

In this appendix, we investigate the effects of differential velocity
aberration on a parallel observation.

The motion of an observer relative to an observed object causes an
apparent displacement of the target on the sky with respect to its
real geometrical position due to velocity aberration. Since the
velocity vector changes during the course of an observation,
corrections must be continuously applied to the telescope tracking
system. These corrections must be computed for a given line-of-sight,
such as the center of the field-of-view of a given instrument in the
case of HST. All the other lines-of-sights will be left partially
uncorrected, leading to what is referred to as differential velocity
aberration.  This effect is much larger for HST than for a ground-based
observation since HST travels at a larger speed.

A first effect of differential velocity aberration is an overall
tracking error of the secondary, parallel instrument, which leads to
an elongated image. A second effect is caused by the difference in
velocity aberration from one corner of the field-of-view of the
instrument to the opposite corner, leading to what appears to be small
change in the image scale over the field of view.

The effect of differential velocity aberration on SPS data can be
investigated by studying what happens in the worst possible case: when
STIS is being used in Parallel Mode while NICMOS is the primary
instrument. This is the worst case since the separation between these
two instruments in the HST focal plane is greater than the one between
STIS and any other instruments on-board HST. Two extreme cases can then
be examined: HST moving in the same direction as the line-of-sight, and
HST moving perpendicularly to the line-of-sight. We assume that HST's
orbit lies in the ecliptic plane and that both the geocentric velocity
of the spacecraft and the velocity of the Earth around the Sun are
constant (7 and 29 km/s respectively).

\begin{figure}
\resizebox{\hsize}{!}{\includegraphics{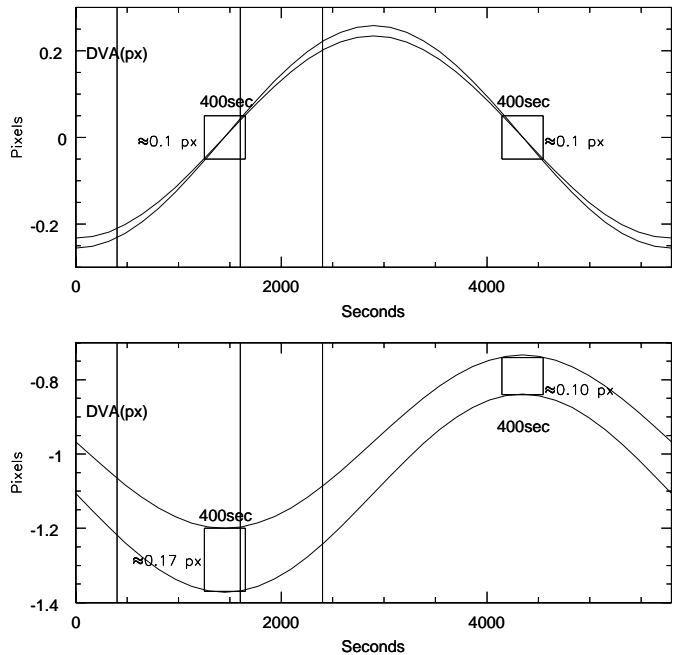}}
\caption{The differential velocity aberration (DVA; in STIS CCD pixel) as a function of
HST orbital period (97 minutes), in two extreme situations (see
text). The two lines show the result for two points laying on opposite
STIS corners. The vertical lines mark some typical STIS Parallel
exposure times. The boxes represent individual 400 second exposures.}
\label{fig:dvasept14}
\end{figure}

The upper panel in Fig. \ref{fig:dvasept14} shows the case of a
line-of-sight that is perpendicular to the direction of the telescope
motion. The lower panel shows the case of the line-of-sight
that is exactly aligned with the velocity vector of the telescope.
We have overlaid three vertical lines to mark time intervals of 400,
1600 and 2400 seconds. These represent typical total exposure
times of individual SPS images and of our STIS Associations containing
4 and 6 consecutive SPS observations.  

On one hand, the amount of shift that can be expected between a set of
successive 400 seconds exposures is given by the amplitude of either
of the two curves in each panel. During the course of a full orbit,
the resulting relative shifts between consecutive images can be as
much as 0.5 STIS CCD pixel ($0.025\arcsec$), if no extra dithering of the primary
instrument is done.  On the other hand, the two curves in each of the
panel of Fig. \ref{fig:dvasept14} show the differential velocity
aberration for two targets located on opposite corners of the STIS
detector. The difference between these two curves therefore provides
an estimate of the image scale variation in a STIS frame, which can be
seen to vary from 0.1 to 0.17 STIS CCD pixel in the case of 400 seconds
exposures. This effect is quite small, but its effect on the
co-addition of images that are offset with respect to one another can
be reduced by restricting the amount of shifts allowed between images
to be co-added, as we have done for this project by limiting the
maximum allowed shifts to be $1/4$ of the field-of-view (Sect.
\ref{sec:selection}).

\end{document}